# Ransomware Threat Mitigation through Network Traffic Analysis and Machine Learning Techniques

Ali Mehrban[1], Shirin Karimi Geransayeh[2]

[1]School of Electrical and Electronic Engineering, Newcastle University, Newcastle, UK
[2]Department of IT Management, Ferdowsi University of Mashhad, Iran

## ABSTRACT

*In recent years, there has been a noticeable increase in cyberattacks using ransomware. Attackers use this malicious software to break into networks and harm computer systems. This has caused significant and lasting damage to various organizations, including government, private companies, and regular users. These attacks often lead to the loss or exposure of sensitive information, disruptions in normal operations, and persistent vulnerabilities. This paper focuses on a method for recognizing and identifying ransomware in computer networks. The approach relies on using machine learning algorithms and analyzing the patterns of network traffic. By collecting and studying this traffic, and then applying machine learning models, we can accurately identify and detect ransomware. The results of implementing this method show that machine learning algorithms can effectively pinpoint ransomware based on network traffic, achieving high levels of precision and accuracy.*

## KEYWORDS

*Keywords: Ransomware, Computer Networks, Cyber Attacks, Network Traffic, Machine Learning*

## 1. INTRODUCTION

Ransomware, a prevalent form of malware, has been proliferating in recent times, posing considerable threats to a wide range of victims, including various organizations and regular users. This menace is not constrained by geographic location or specific operating systems. According to reports from "Cyber Security Ventures", the total damage caused by ransomware to organizations and individuals in 2019 amounted to approximately $11.5 billion. Shockingly, every 14 seconds, a user or organization falls victim to a ransomware attack, a timeframe that has reduced to 11 seconds in 2021[1]. Within this remarkably short span, cyberattacks through ransomware have escalated, becoming increasingly perilous and invasive.

These attacks have impacted diverse sectors, including finance, insurance, banking, real estate, healthcare, and government management. Symantec Security reports a 46% increase in the number of ransomware types in 2017 [2]. Ransomware restricts victims' access to their files by encrypting targeted files until the ransom is paid [3]. In 2017, the "No More Ransom" project was initiated as a collaboration between European law enforcement and information technology security companies, aiming to disrupt criminal activities associated with ransomware and assist businesses and individuals in mitigating its impact [4]. Similar commercial software products have also been developed for network defense.

Security solutions like "Cybereason" utilize behavioral techniques to protect consumer networks [5], while "Darktrace" employs advanced unsupervised machine learning techniques for organizational network protection [6]. Several machine learning techniques and frameworks for ransomware detection and identification have been proposed and implemented. Malware analysis and detection have been subjects of research for years, involving both static and dynamic analysis approaches, with various analysis tools recommended [7], [8]. Additionally, various classification approaches for malware have been suggested [9], [10], but these approaches may not be suitable for defending against ransomware, as they generally focus on distinguishing malware from benign software.

Therefore, specialized detection mechanisms for ransomware are necessary, focusing on specific ransomware features to differentiate it from other malware types and benign software. This article proposes a method for detecting and identifying ransomware in computer networks using machine learning techniques.

The steps for ransomware detection and identification through network traffic analysis are as follows:

1- Study, evaluation, and analysis of network traffic for 54 ransomware families with 396 samples and 5 benign software types with 420 samples.

2- Evaluation of machine learning algorithms for ransomware detection and identification based on network traffic.

3- Extraction and utilization of ransomware-specific features from network traffic using the T-shark network protocol analyzer [11].

This article is organized as follows: Section 2 discusses relevant works on machine learning, ransomware detection and identification, and network traffic analysis. Section 3 outlines the proposed method. Section 4 examines and evaluates the proposed method. Finally, Section 5 presents the conclusions.

## 2. RELATED WORK

In recent years, there have been many techniques for detecting and identifying ransomware using a combination of machine learning and deep learning with the development of artificial intelligence. Various ransomwares can be analyzed with a fully defined behavioral structure, and most ransomware families have common behavioral features including payload persistence, obfuscation techniques, and network traffic. Therefore, the related articles are as follows:

In 2022, Masum and colleagues presented a feature selection-based framework using various machine learning algorithms, including neural network-based architectures, for security level classification for ransomware detection and prevention [12]. They also used several machine learning algorithms in this article, including Random Forest, Decision Tree, Logistic Regression, Naïve Bayes, and Neural Network, which were used on a number of selected features for ransomware classification. All experiments were conducted on a ransomware dataset to evaluate the proposed framework. Experimental results show that the Random Forest algorithm performs better than other algorithms in terms of accuracy, F-beta, and precision scores.

In 2020, Ganta and colleagues proposed an approach using machine learning, which is unlike the traditional ransomware detection system [13]. This framework used various classification algorithms including Logistic regression, Decision tree, Random forest, and KNN algorithm for detecting hidden ransomware in executable files. In 2018, Poudyal and colleagues introduced a machine learning-based detection model for efficient ransomware identification, which uses multi-level analysis for better interpretation of the target sections of malware code [14]. The model was evaluated and the results indicate its performance in ransomware identification between 76% to 97%. In 2015, Cabaj and colleagues conducted a detailed analysis of the CryptoWall ransomware [15], but they only analyzed network activities and found that CryptoWall uses the HTTP protocol for communications and some obfuscation techniques such as Tor and fake DNS requests have been used in this attack.

In 2018, Wan and colleagues proposed a network intrusion detection framework consisting of Argus server and client applications, introducing a new flow-oriented method as Biflow for ransomware detection [16]. For classifying datasets, six algorithms were used and supervised machine learning was used to achieve better accuracy and increase the performance of the detection module. The RandomForest algorithm is one of the popular machine learning algorithms in this method, which is used for malware and ransomware detection. To improve common approaches, the concept of advanced machine learning should be used in ransomware detection and prevention. In 2018, Kim evaluated the effectiveness of using machine learning techniques in malware detection and he extracted the features of Windows system calls and used 10-grams, 9-grams, 8-grams to extract feature information. Finally, he implemented his work with SVM [34] and SGD algorithms and achieved 96% detection accuracy [17].

In 2018, Abiola and colleagues presented a signature-based detection model for malware, extracting Brontok worms, and for signature parsing, an n-gram technique was used [18]. Signature-based analysis is the most commonly used traditional anti-malware system. In 2020, Noorbehbahani and colleagues focused on semi-supervised learning to take advantage of some labeled data and many unlabeled data for ransomware identification [19]. Various feature selection and semi-supervised classification methods were used in the CICandMal 2017 datasets for ransomware analysis, and the semi-supervised classification method using Random Forest as the base classifier performs better than various semi-supervised classification techniques for ransomware identification. In 2020, Khan and colleagues proposed a ransomware detection framework based on the DNAact-Ran digital DNA sequence engine, which focuses on the limitations of sequence design and k-mer frequency vector [20]. This framework was demonstrated on 582 ransomware-DNAact Run and 942 benign software for measuring the performance of recall, precision, and f-measure.

In 2017, Maniath and colleagues proposed a framework for classifying binary sequences of API calls using Long Short-Term Memory (LSTM) networks for ransomware classification through its behavior [21]. A dynamic analysis technique was used to extract API calls from modified system input in a sandbox environment. According to the evaluation, the proposed LSTM-based framework achieved 96.67% accuracy in automatic ransomware behavior classification from a large volume of malware datasets. However, by strengthening the LSTM network, the overall accuracy can be further improved.

In 2018, Alhawi and colleagues introduced the NetConverse model, a machine learning analysis of Windows ransomware network traffic, to achieve a high and stable detection rate using a dataset created from conversation-based network traffic features, using the Decision Tree algorithm to achieve a true positive detection rate of 97.1% [22].

In 2019, Masum and colleagues proposed a deep learning framework (Droid-NNet) for malware classification in Android, which is a mechanized deep learner that performs better than existing advanced machine learning methods [23]. Based on the evaluation report of two Android application datasets, Malgenome-215 and Droid-NNet, Drebin-215 shows strong and effective malware detection on the Android platform. The proposed framework was optimized with a limited number of important features and tested with various machine learning classifiers, including neural network-based architectures. Experimental results demonstrate the robustness and effectiveness of the proposed framework. In 2015, Damshenas and colleagues presented a method that used behavioral features of malware such as API calls, system file changes, and network traffic as features for various classification tasks, therefore, machine learning techniques can be used for ransomware detection [24].

## 3. METHODOLOGY

The proposed method of this research is to detect and identify ransomware through network traffic using machine learning algorithms. In general, this proposed method includes the following two sections:

### 3.1. Training and Creating a Machine Learning Model:

In the first part of this proposed method, samples of network traffic containing significant ransomware and benign software are initially collected. Then, feature extraction and selection are performed, and these extracted and selected features are used as input for training machine learning algorithms and converting them into a machine learning model for detecting and identifying ransomware.

**3.1.1. Data Collection:** In the first stage of this section, samples of both types of network traffic containing significant ransomware and benign software are collected and merged to create a dataset. Data collection and preprocessing is an important step in training machine learning algorithms.

**3.1.1.1: Ransomware Used:** The network traffic of the ransomware that has been implemented is collected through the website Ransomware Tracker [25] and articles [26] and [27]. In total, it includes 396 samples of ransomware network traffic. It includes the following items:

| Ransomware Family | Sample Size | Ransomware Family | Sample Size | Ransomware Family | Sample Size |
|---|---|---|---|---|---|
| Aleta | 1 | Shade | 4 | Scarab | 1 |
| Bart | 1 | DMALocker | 1 | Sodinokibi | 6 |
| BitPaymer | 1 | Eris | 1 | TeslaCrypt | 28 |
| Cerber | 45 | GandCrab | 2 | Paycrypt | 2 |
| CryLock | 1 | Hive | 1 | Padcrypt | 1 |
| CrypMIC | 1 | GlobeImposter | 1 | Cryptolocker | 30 |
| CryptFile2 | 1 | Jaff | 2 | Cryptowall | 30 |
| CryptoFortress | 1 | Locky | 40 | Torrentlocker | 30 |
| CryptoMix | 4 | MRCR | 1 | WannaLocker | 10 |
| CryptoShield | 2 | Maktub | 1 | Simplocker | 10 |
| Crysis | 12 | Maze | 1 | RansomBO | 10 |
| Phobos | 1 | Mole | 1 | PornDroid | 10 |
| Razi | 1 | Netwalker | 1 | Pletor | 10 |
| Ryuk | 2 | RansomX | 1 | LockerPin | 10 |
| Sage | 2 | Revenge | 1 | Koler | 10 |
| Spora | 1 | Stop | 2 | Jisut | 10 |
| Zeus | 1 | Svpeng | 11 | Charger | 10 |
| WannaCry | 4 | Virlock | 3 | CTB-Locker | 31 |

Table 1: Ransomware families investigated in the implementation

**3.1.1.2. Benign Software Used:** Network traffic samples of benign software have also been collected from the Virus Total Intelligence platform [28]. The criteria for selecting samples of benign software network traffic are that they have passed through the network at least 3 times and have not been identified as ransomware by antiviruses and firewalls. In total, 420 samples of benign software network traffic have been collected.

**3.1.2 Feature Extraction and Selection:** In the second phase of this section, feature extraction from network traffic is performed using TShark. The features extracted from network traffic include static features and important statistical data, which are based on 5 criteria [29]: protocol, source/destination IP address, source/destination port values, which are aggregated in unique files, and equivalent statistical values are also extracted. The extracted and selected features from network traffic are merged with each other to create preprocessed data.

| Feature name | Description |
|---|---|
| Address A | Source host IP address |
| Port A | Source host port number |
| Address B | Destination host IP address |
| Port B | Destination host port number |
| Protocol | Protocol type |
| Relative start | Time relative to the start of the conversation (in seconds) |
| Duration | Conversation duration (in seconds) |
| Packets A → B | Number of packets sent from A to B |
| Bytes A → B | Number of bytes sent from A to B |
| Packets | Total number of packets in each conversation |
| Bytes | Total number of bytes in each conversation |
| Packets B → A | Number of packets sent from B to A |
| Bytes B → A | Number of bytes sent from B to A |

Table 2: Ransomware families investigated in the implementation

### 3.1.3 Training and Building a Machine Learning Model

In the third step of this section, the extracted and selected features from network traffic are given as input to machine learning algorithms for training and building a machine learning model that can be used to detect and identify ransomware through network traffic in the second section.

### 3.2. Machine Learning Model Testing Section

In the second section of the proposed method for detecting and identifying ransomware through network traffic, the machine learning model trained in the first section is installed on one of the network devices and a copy of the network traffic containing ransomware and benign software is sent to this device in addition to the destination, and then the following steps are performed.

### 3.2.1. Data Collection

In the first step of this section, network traffic containing ransomware and benign software is collected at pre-defined time intervals. The following table shows a sample of the collected data set:

| Protocol | Address A | Port A | Address B | Port B | Packets | Bytes | Packets A → B | Bytes A → B | Packets B → A | Bytes B → A | Rel Start | Duration |
|---|---|---|---|---|---|---|---|---|---|---|---|---|
| 6 | 192.168.1.4 | 49252 | 192.168.1.5 | 5357 | 20 | 15137 | 8 | 1396 | 12 | 13741 | 1.841135 | 0.026054 |
| 6 | 104.20.18.218 | 443 | 10.0.2.4 | 49334 | 10 | 845 | 10 | 845 | 3 | 256 | 13.92507 | 481.2216 |
| 6 | 192.168.56.10 | 58768 | 216.58.54.7 | 80 | 4 | 1537 | 59 | 11327 | 18 | 3455 | 183.9047 | 93.32879 |
| 6 | 192.168.56.14 | 49273 | 35.161.88.115 | 43 | 13 | 4509 | 15 | 1642 | 28 | 6151 | 18.90573 | 61.66712 |
| 17 | 169.254.49.74 | 59463 | 224.0.0.252 | 5355 | 2 | 126 | 2 | 126 | 15 | 945 | 21.09118 | 0.127367 |
| 17 | 192.168.1.5 | 3702 | 192.168.1.4 | 51167 | 2 | 2536 | 2 | 2536 | 28 | 35504 | 3.103964 | 0.062291 |
| 17 | 192.168.38.17 | 65819 | 37.185.88.2 | 443 | 12 | 4276 | 1 | 590 | 4 | 2360 | 97.64251 | 0.781341 |

Table 3: Samples of Dataset

### 3.2.2 Feature extraction and selection

In the second step of this section, the process of extracting and selecting important and useful features from network traffic for testing the machine learning model for detecting and identifying ransomware from among benign software is performed, as in the second step of the first section. The extracted and selected features from network traffic are the same as Table 2.

### 3.2.3 Testing the machine learning model

In the third step of this section, the important extracted and selected features from network traffic are given as input to the machine learning model trained in the first section that is placed on one of the network devices. After processing by the machine learning model, if ransomware passes through the network traffic, a warning message is issued that ransomware has been detected and identified.

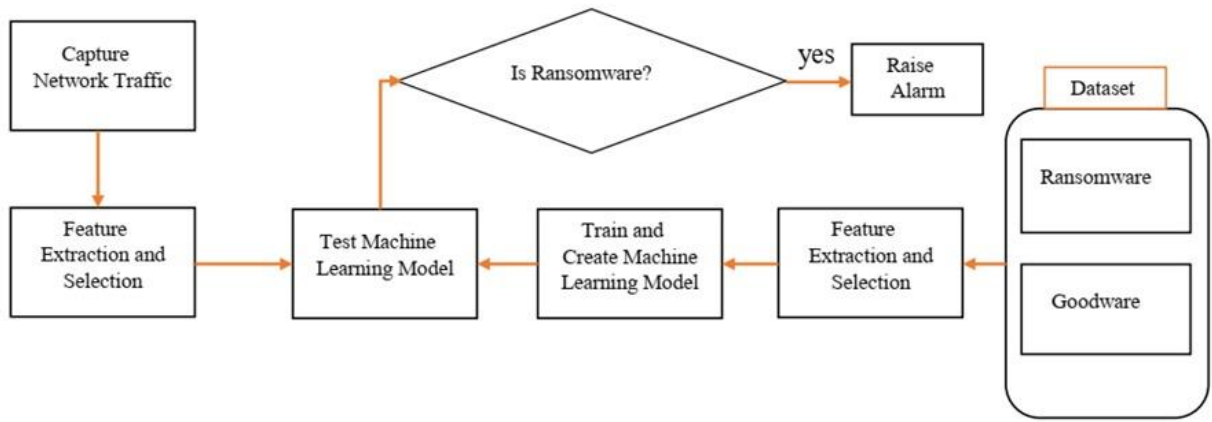

Figure 1: General overview of the proposed method

## 4. Review and Evaluation of the Proposed Method:

In this section, the review and evaluation of the proposed method for achieving the highest rate of ransomware detection and identification with high precision and accuracy have been addressed. The proposed method is designed in such a way that, using 13 important features extracted and selected from network traffic, machine learning algorithms have been trained and tested.

### 4.1 Review of Machine Learning Algorithms:

The review of machine learning algorithms based on their advantages and disadvantages is as follows:

| Algorithm | Pros | Cons |
|---|---|---|
| K-Nearest Neighbor | Simple and requires no training | High processing time |
| Multilayer Perceptron | Accurate estimation | High processing time |
| J48 | Fast classification and scalability | Inaccurate for continuous class values and features |
| Random Forest | Ability to improve prediction performance | Hard to analyze output |
| Support Vector Machines | Relatively efficient memory | Not suitable for large datasets |
| Bayes Network | Fast training | Impractical for large datasets |

Table 4: Machine Learning Algorithms

## 4.2 Evaluation Measures for Ransomware Detection and Identification Rates in Machine Learning Algorithms:

Evaluation criteria are important indicators used as performance metrics for a predictive model based on machine learning approaches. The performance of ransomware detection and identification rates for machine learning algorithms has been evaluated using 6 standard criteria from the Scikit-Learn machine learning library.

| Evaluation metric | Calculation method | Description |
|---|---|---|
| True positive rate (TPR) | TP / (TP + FN) | The percentage of ransomware that is correctly classified. |
| False positive rate (FPR) | FP / (FP + TN) | The percentage of benign software that is incorrectly classified as ransomware. |
| Precision | TP / (TP + FP) | The percentage of ransomware that is predicted correctly. |
| Recall | TPR | The percentage of all ransomware that is predicted correctly. |
| F-measure | 2 × (Recall × Precision) / (Recall + Precision) | A measure of the overall performance of the system. |
| Accuracy score | (TP + TN) / (TP + FN + TN + FP) | The percentage of all samples that are predicted correctly. |

Table 5: Evaluation Criteria (TP= True positive, FN=False negative, TN= True negative, FP=False positive)

## 4.3 Evaluation Results of the Proposed Method

The following table shows the processing time spent on building the machine learning models used in the implementation.

| Classifier | Processing time(in seconds) |
|---|---|
| K-Nearest Neighbor | 0.09 |
| Multilayer Perceptron | 461.5 |
| J48 | 9.36 |
| Random Forest | 52.17 |
| Support Vector Machines | 273.8 |
| Bayes Network | 7.63 |

Table 6: Comparison of Processing Time (in Seconds)

The following table shows a comparison of the evaluation metrics for the ransomware detection and identification rates for machine learning algorithms.

| Classifier | TPR(%) | FPR(%) | Precision | Recall | F-measure | Accuracy_score |
|---|---|---|---|---|---|---|
| K-Nearest Neighbor | 92.80 | 6.70 | 93.10 | 92.80 | 92.90 | 92.90 |
| Multilayer Perceptron | 97.30 | 1.80 | 97.50 | 97.30 | 97.40 | 97.60 |
| J48 | 94.20 | 5.30 | 94.20 | 94.20 | 94.20 | 94.00 |
| Random Forest | 95.70 | 4.50 | 95.80 | 95.70 | 95.70 | 96.10 |
| Support Vector Machines | 93.40 | 3.60 | 93.60 | 93.40 | 93.50 | 93.70 |
| Bayes Network | 95.60 | 4.60 | 95.90 | 95.60 | 95.70 | 95.80 |

Table 7 Comparison of Criteria for Evaluating Detection and Detection Rate of Ransomware

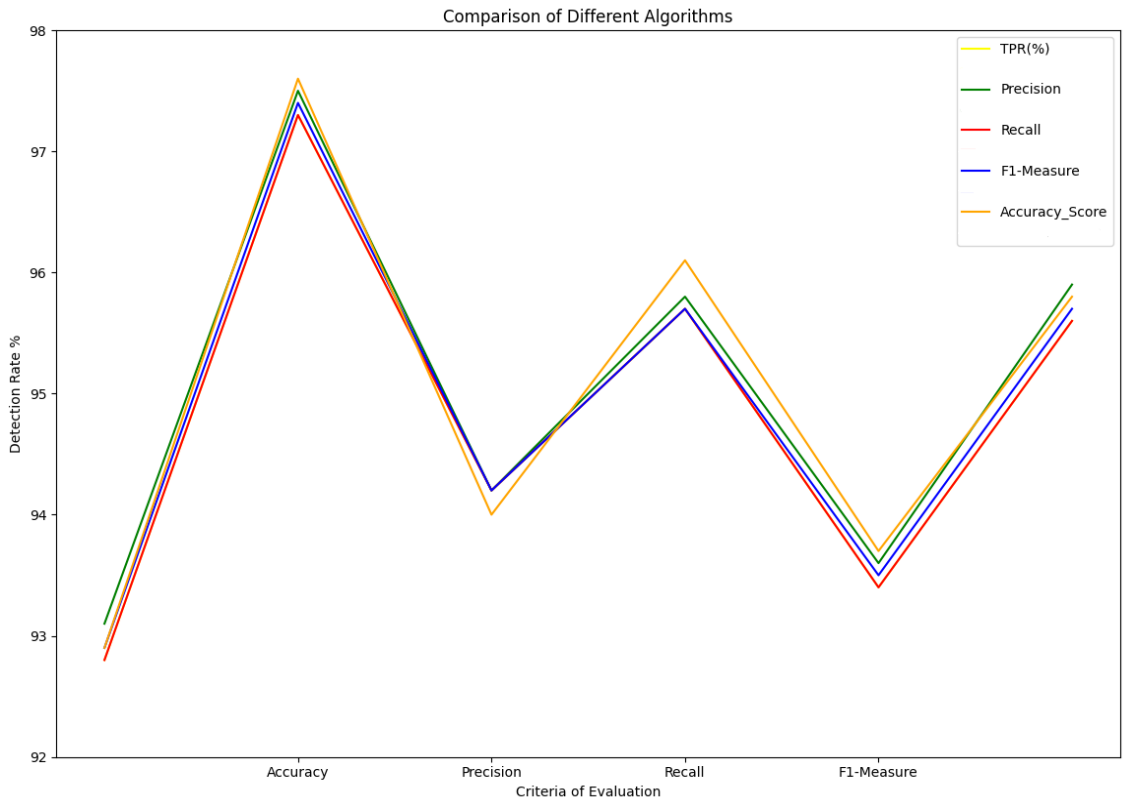

Figure 2: Criteria of Evaluation vs Detection Rate

### 4.4 Comparison of Evaluation Results of Similar Studies

In a study and evaluation conducted to detect Android malware [30], using a dataset created by extracting network traffic features, especially statistical features, it was concluded that with machine learning techniques using the Random Forest algorithm, it is possible to reach 99.9% TPR. In comparison with the evaluation performed by the proposed method using the Multilayer Perceptron machine learning algorithm, 97.30% TPR was achieved. In another study and evaluation conducted to detect Windows ransomware [31], using the extraction of network traffic features, through dynamic analysis, 96.3% TPR was achieved. In detecting botnets by tracking conversations [32], [33], [35] a statistical network conversation approach was used to analyze botnet traffic and achieved an average of 95.0% TPR in detecting 4 different botnet programs.

| Study | Accuracy (TPR %) |
|---|---|
| NetConverse [4] | 97.1 |
| Peershark [19] | 95.0 (average) |
| EldeRan [8] | 96.3 |
| Mobile malware detection [12] | 99.99 |
| Proposed Method | 97.3 |

Table 8: Comparison of the Results of Similar Studies

## 5. CONCLUSION

This research presented a proposed method for detecting and identifying ransomware based on network traffic using machine learning algorithms. It showed a high accuracy rate in correctly detecting and identifying ransomware from among benign software. In this proposed method, network traffic is analyzed using machine learning algorithms such as Random Forest, J48, Multilayer Perceptron, K-Nearest Neighbor, Support Vector Machines, Forest, and Bayes Network. If the network traffic is suspicious of ransomware, an alert is issued that ransomware has been detected and identified. The advantages of this proposed method over previous methods are that in this proposed method, more samples of ransomware and benign software have been reviewed, and also the appropriate setting and initialization have been carried out in relation to the parameters used in machine learning models, which has led to an increase in accuracy in detecting and identifying ransomware. The results of implementing the proposed method indicate high performance along with high accuracy of machine learning algorithms for detecting and identifying ransomware. The findings and results of this proposed method show an accuracy rate of 97.30% TPR with the Multilayer Perceptron machine learning algorithm. This method shows that machine learning algorithms can detect and identify ransomware based on features extracted from network traffic.

## REFERENCES

*1. Usharani, S.; Bala, P.M.; Mary, M.M.J. Dynamic analysis on crypto-ransomware by using machine learning: Gandcrab ransomware. J. Phys. Conf. Ser. 2021, 1717, 012024.*

*2. Wood P, Egan G. Symantec Internet Security Threat Report 2011. Technical report. Mountain View, CA: Symantec Corp; 2012.*

*3. Zakaria WZA, Abdollah MF, Mohd O, Ariffin AFM. The rise of ransomware. In: Proceedings of the 2017 International Conference on Software and e-Business (ICSEB); 2017; Hong Kong.*

*4. "The No More Ransom Project." [Online]. Available: https://www.nomoreransom.org/. [Accessed: 31-Mar-2017].*

*5."Ransomware Protection – Ransom free by Cybereason." [Online]. Available: https://ransomfree.cybereason.com/. [Accessed: 31-Mar-2017].*

*6."Darktrace Technology" [Online]. Available: https://www.darktrace.com/technology/#machine-learning. [Accessed: 31-Mar-2017].*

*7. Arends J, Kerstin I. Malware analysis: tools and techniques. Sankt Augustin, Germany: University of Applied Science Bonn-Rhein-Sieg; 2018.*

*8. Mehrban, Ali, and Pegah Ahadian. "Malware Detection in IOT Systems Using Machine Learning Techniques." arXiv preprint arXiv:2312.17683 (2023).*

*9. Obeis NT, Bhaya W. Malware analysis using APIs pattern mining. Int J Eng Technol. 2018;7(3.20):502-506.*

*10. Sami A, Yadegari B, Rahimi H, Peiravian N, Hashemi S, Hamze A. Malware detection based on mining API calls. In: Proceedings of the 2010 ACM Symposium on Applied Computing (SAC); 2010; Sierre, Switzerland.*

*11. "tshark\ -\ The\ Wireshark\ Network\ Analyzer\ 2.0.0," 2017. [Online]. Available: https://www.wireshark.org/doc/man-pages/tshark.html. [Accessed: 29-May-2017].*

*12. M. Masum, M. J. Hossain Faruk, H. Shahriar, K. Qian, D. Lo and M. I. Adnan, "Ransomware Classification and Detection with Machine Learning Algorithms," 2022 IEEE 12th Annual Computing and Communication Workshop and Conference (CCWC), Las Vegas, NV, USA, 2022, pp. 0316-0322, doi: 10.1109/CCWC54503.2022.9720869.*

*13. V. G. Ganta, G. V. Harish, V. P. Kumar and G. R. K. Rao, "Ransomware Detection in Executable Files Using Machine Learning," 2020 International Conference on Recent Trends on Electronics, Information, Communic ation & Technology (RTEICT), Bangalore, India, 2020, pp. 282-286, doi: 10.1109/RTEICT49044.2020.9315672.*

*14. S. Poudyal, K. P. Subedi and D. Dasgupta, "A Framework for Analyzing Ransomware using Machine Learning," 2018 IEEE Symposium Series on Computational Intelligence (SSCI), Bangalore, India, 2018, pp. 1692-1699, doi:10.1109/SSCI.2018.8628743.*

*15. Cabaj K, Gawkowski P, Grochowski K, Osojca D. Network activity analysis of CryptoWall ransomware. Przeglad Elektrotechniczny. 2015;91(11):201-204.*

*16. Y. -L. Wan, J. -C. Chang, R. -J. Chen and S. -J. Wang, "Feature-Selection-Based Ransomware Detection with Machine Learning of Data Analysis," 2018 3rd International Conference on Computer and Communication Systems (ICCCS), Nagoya, Japan, 2018, pp. 85-88, doi:10.1109/CCOMS.2018.8463300.
17. Kim CW. NtMalDetect: a machine learning approach to malware detection using native API system calls. arXiv preprint arXiv:1802.05412. 2018.*

*18. Abiola, Alogba & Marhusin, M.F. (2018). Signature-Based Malware Detection Using Sequences of N-grams. International Journal of Engineering and Technology(UAE). 7. 10.14419/ijet.v7i4.15.21432.*

*19. F.Noorbehbahani and M. Saberi, "Ransomware Detection with Semi-Supervised Learning,"* 2020 10th International Conference on Computer and Knowledge Engineering (ICCKE), *Mashhad, Iran, 2020, pp. 024-029, doi:10.1109/ICCKE50421.2020.9303689.*

*20. F. Khan, C. Ncube, L. K. Ramasamy, S. Kadry and Y. Nam, "A Digital DNA Sequencing Engine for Ransomware Detection Using Machine Learning," in IEEE Access, vol. 8, pp. 119710-119719, 2020, doi: 10.1109/ACCESS.2020.3003785.*

*21. S. Maniath, A. Ashok, P. Poornachandran, V. G. Sujadevi, P. Sankar A.U. and S. Jan, "Deep learning LSTM based ransomware detection," 2017 Recent Developments in Control, Automation & Power Engineering (RDCAPE), Noida, India, 2017, pp. 442-446, doi: 10.1109/RDCAPE.2017.8358312.*

*22. Alhawi, O.M.K., Baldwin, J., Dehghantanha, A. (2018). Leveraging Machine Learning Techniques for Windows Ransomware Network Traffic Detection. In: Dehghantanha, A., Conti, M., Dargahi, T. (eds) Cyber Threat Intelligence. Advances in Information Security, vol 70. Springer, Cham. https://doi.org/10.1007/978-3-319-73951-9_5*

*23. M. Masum and H. Shahriar, "Droid-NNet: Deep Learning Neural Network for Android Malware Detection," 2019 IEEE International Conference on Big Data (Big Data), Los Angeles, CA, USA, 2019, pp. 5789-5793, doi:10.1109/BigData47090.2019.9006053.*

*24. M. Damshenas, A. Dehghantanha, K.-K. R. Choo, and R. Mahmud, "M0Droid: An Android Behavioral-Based Malware Detection Model," J. Inf. Priv. Secur., vol. 11, no. 3, pp. 141–157, Jul. 2015.*

25. "Tracker|RansomwareTracker," 2016. [Online]. Available: https://ransomwaretracker.abuse.ch/tracker [Accessed : 04-Jan-2017].

26. A. H. Lashkari, A. F. A. Kadir, L. Taheri and A. A. Ghorbani, "Toward Developing a Systematic Approach to Generate Benchmark Android Malware Datasets and Classification," 2018 International Carnahan Conference on Security Technology (ICCST), Montreal, QC, Canada, 2018, pp. 1-7, doi: 10.1109/CCST.2018.8585560.

27. E. Berrueta, D. Morato, E. Magaña and M. Izal, "Open Repository for the Evaluation of Ransomware Detection Tools," in IEEE Access, vol. 8, pp. 65658-65669, 2020, doi: 10.1109/ACCESS.2020.2984187.

28. "Virus Total - Free Online Virus, Malware and URL Scanner." [Online]. Available: https://www.virustotal.com/. [Accessed: 31-Mar-2017].

29. P. Narang, C. Hota, and V. Venkatakrishnan, "PeerShark: flow-clustering and conversation-generation for malicious peer-to-peer traffic identification," EURASIP J. Inf. Secur., vol. 2014, no. 1, p. 15, 2014.

30. F. A. Narudin, A. Feizollah, N. B. Anuar, and A. Gani, "Evaluation of machine learning classifiers for mobile malware detection," Soft Comput., vol. 20, no. 1, pp. 343–357, 2016.

31. D. Sgandurra, L. Muñoz-González, R. Mohsen, and E. C. Lupu, "Automated Dynamic Analysis of Ransomware:Benefits, Limitations and use for Detection," no. September, 2016.

32. P. Narang, S. Ray, C. Hota, and V. Venkatakrishnan, "PeerShark: Detecting peer-to- peer botnets by tracking conversations," in Proceedings - IEEE Symposium on Security and Privacy, 2014, vol. 2014–Janua, pp. 108-11

33. Mehrban, Ali, and Pegah Ahadian. "An adaptive network-based approach for advanced forecasting of cryptocurrency values." arXiv preprint arXiv:2401.05441 (2024).

34. P. Ahadian and K. Parand, "Support vector regression for the temperature-stimulated drug release," Chaos, Solitons & Fractals, vol. 165, p. 112871, 2022.

35. Mehrban, Ali, and Pegah Ahadian. "Evaluating BERT and ParsBERT for Analyzing Persian Advertisement Data." International Journal on Natural Language Computing (IJNLC) Vol 12 (2023).